\documentclass[11pt,twocolumn]{article}
\usepackage{graphics}
\usepackage{graphics}

\newcommand{\cyan}{CH$_3$CN }
\newcommand{\meth}{CH$_3$OH } 

\makeatletter
\renewcommand{\@biblabel}[1]{}
\makeatother
\title{\begin{flushleft}
\bf Determination of molecular gas properties using methyl cyanide
lines
\end{flushleft}}
\author{
\small\bf S.V. Kalenskii$^1$, V.G. Promislov$^1$, A.V. Alakoz$^1$,
A. Winnberg$^2$, and L.E.B. Johansson$^2$~~~~~~~~\\
\small\rm $^1$~Astro Space Center of Lebedev Physical Institute,
Profsoyuznaya 84/32, 117810 Moscow, Russia~~~~~~~~~~~~\\
\small\rm email: kalensky@dpc.asc.rssi.ru, vitaly@diogen.asc.rssi.ru,
alakoz@dpc.asc.rssi.ru~~~~~~~~~~~~~~~~~~~~~~~~~~~~`~\\
\small\rm
$^2$~Onsala Space Observatory, S-439 92, Onsala, Sweden~~~~~~~~
~~~~~~~~~~~~~~~~~~~~~~~~~~~~~~~~~~~~~~~~~~~~~~~~~~~~~~~~~~~~~\\
\small\rm
e-mail: anders@oso.chalmers.se, leb@oso.chalmers.se~~~~~~~~~~~~~~~~~~~~~~
~~~~~~~~~~~~~~~~~~~~~~~~~~~~~~~~~~~~~~~~~~~~
}

\begin{document}
\maketitle

\begin{abstract}
\small
A survey of 27 galactic star-forming regions in the $6_K-5_K$, $5_K-4_K$,
and $8_K-7_K$ \cyan lines at 110, 92, and 147~GHz, respectively, was made.
Twenty-five sources were detected at 110 GHz, nineteen at 92 GHz, and
three at 147~GHz. The strongest \cyan emission arise
in hot cores in the regions of massive star formation. \cyan abundance
in these objects is larger than $10^{-9}$ due to grain
mantle evaporation. Weaker \cyan lines were found in a number of
sources. They may arise either in warm \mbox{(30-50~K)} dense
($10^5$--$10^7$~cm$^{-3}$) clouds, or in hot regions accompanied by colder 
gas. 
\end{abstract}


\subsection*{1. Introduction}

Methyl cyanide (CH$_3$CN) radio emission was first detected in space by
Solomon et al. (1971). Like other symmetric top molecules, e.g., ammonia or
methyl acetylene, methyl cyanide is a good temperature probe for
interstellar gas. Since radiative transitions between different
$K$-ladders are prohibited in symmetric tops, the population of different
$K$-ladders is determined by collisions and relative intensities 
of lines with different quantum numbers $K$ are strongly related to kinetic
temperature. \cyan has a large number of rotational transitions, which
are grouped in series of closely spaced lines with the same $J$ but different
$K$ quantum numbers. These lines can be observed simultaneously with the same
receiver and spectrometer. Therefore the kinetic temperature estimates
are not affected by calibration and pointing errors (Hollis et al. 1981; 
Loren \& Mundy 1984). The \cyan spectral characteristics are well known 
(see e.g. Boucher et al. 1980; Loren \& Mundy 1984).

With its high dipole moment (3.91 debyes) \cyan is a tracer of dense gas. It
is well known that the \cyan radio emission often arises in hot gas, in
particular in hot cores, i.e. hot and dense neutral regions with sizes a few
0.01 pc. Interferometric observations of several strong \cyan sources show 
that the hot core emission in their spectra is dominant, but sometimes is 
accompanied by a much weaker halo emission (see, e.g., Olmi et al. 1996, 
and references therein). The reason for strong \cyan emission in hot cores 
is grain mantle evaporation, which leads to a significant increase of \cyan 
abundance. This may occur either due to \cyan evaporation or due to a
chain of gas-phase reactions, starting from HCN or some other
nitrogen-bearing molecule, which appears in gas phase due to mantle
evaporation (Millar 1996). Besides hot cores, \cyan has been detected 
toward warm clouds (30--50~K) in regions of massive star formation 
(Olmi et al. 1993) and in dark clouds (Irvine~et~al.~1987). 
In spite of a relatively large number of interferometric observations
of hot cores in various \cyan lines, only a few \cyan surveys have been
published, each of them devoted only to a small number of objects (Bergman \&
Hjalmarson 1989; Churchwell et al. 1992; Olmi et al. 1993). Therefore,
following our previous studies of molecular clouds in various
methanol lines (Kalenskii~et~al.~1997; Slysh~et~al.~1999), we made a
more extended survey of Galactic star-forming regions in the $6_K-5_K$ and
$5_K-4_K$ series of methyl cyanide lines near 110 and 92~GHz, respectively.
Three sources were additionally observed in the $8_K-7_K$ lines near
147~GHz.

\begin{table}[t]
\caption{Frequencies and line strengths of the \cyan transitions
\label{lines}}
\[
\begin{array}{lrr}
\hline\noalign{\smallskip}
{\rm Transition}\rule{0pt}{11pt}\qquad
               &\qquad{\rm Frequency^1}
                           &{\rm Line}\\
               &{\rm (MHz)}\quad &\qquad {\rm strength}\\
\noalign{\smallskip}               
\hline\noalign{\smallskip}
5_0-4_0     &91987.090  &5.00\\
5_1-4_1     &91985.317  &4.80\\
5_2-4_2     &91980.089  &4.20\\
5_3-4_3     &91971.310  &3.20\\
5_4-4_4     &91959.024  &1.80\\
6_0-5_0     &110383.522 &6.00\\
6_1-5_1     &110381.404 &5.83\\
6_2-5_2     &110375.052 &5.33\\
6_3-5_3     &110364.469 &4.50\\
6_4-5_4     &110349.659 &3.33\\
6_5-5_5     &110330.627 &1.83\\
8_0-7_0     &147174.592 &8.00\\
8_1-7_1     &147171.756 &7.88\\
8_2-7_2     &147163.248 &7.50\\
8_3-7_3     &147149.073 &6.88\\
8_4-7_4     &147129.236 &6.00\\
8_5-7_5     &147103.744 &4.87\\
\noalign{\smallskip}
\hline\noalign{\smallskip}
\end{array}
\]
{\footnotesize $^1$--from the database by Lovas (http://physics.nist.gov/)}
\end{table}

\subsection*{2. Observations}

The observations at 92 and 110~GHz were carried out in April--May 1996
using the 20-m millimetre-wave telescope of the Onsala Space Observatory.
The pointing accuracy
was checked by observations of SiO masers and was found to be within 5$''$.
Main beam efficiency and half-power beamwidth at 92~GHz were 0.55 and $39''$,
respectively. The observations were performed in a dual-beam switching mode
with a  switch frequency of 2 Hz and a beam separation of 11$'$. 
In no case we detected absorption features indicating that
the OFF beam crossed another source component.
A cryogenically cooled low-noise SIS mixer was used at both frequencies.  The
single sideband receiver noise temperature was about 150~K. The system noise
temperature, corrected for atmospheric absorption, rearward spillover and
radome losses, varied at both frequencies between 350 and 2000~K, depending
on the weather conditions and source elevation. The data were calibrated
using the standard chopper-wheel method. The backend consisted of two
parallel filter spectrometers: a 256-channel spectrometer with 250 kHz
frequency resolution and a 512-channel spectrometer with 1~MHz frequency
resolution. Two objects, NGC~1333~IRAS2 and L~1157, were observed only with
the 1~MHz resolution.

The observations at 147~GHz were carried out on October 4, 1998
using the 12-m telescope NRAO\footnote[1]{NRAO is operated by Associated
Universities, Inc., under contract with the National Science Foundation}
at Kitt-Peak, AZ, in a remote-observing mode from the Astro Space Center,
Moscow. The observations were performed in a position switching mode
with a position separation of 5$'$. A cooled SIS-mixer, dual polarization
channel receiver was used. Both senses of polarization were observed 
simultaneously. The data were calibrated using the standard vane method. 
The system noise temperature, corrected for atmospheric absorption, rearward 
spillover and ohmic losses, varied between 300 and 500~K. The pointing 
accuracy was found to be within 5$''$. Main beam efficiency and half-power 
beamwidth were 0.54 and $42''$, respectively. The backend consisted of two 
512-channel filter bank spectrometers with 1~MHz frequency resolution, 
connected to different polarization channels.

The frequencies and strengths of the observed \cyan lines
are presented in Table~\protect\ref{lines}. 

At 110 GHz 27 sources were observed. Of those 19 were observed at 92 GHz
and 3 at 147~GHz. Spectra were reduced using the Grenoble CLASS package. 
We made Gaussian fitting assuming that different $K$-components in each series
have common LSR velocities and linewidths. The results are presented in 
Table~\protect\ref{gparm}.

Five objects, Ori~S6, G30.8-0.1, G34.26+0.15, W~51E1/E2 and DR~21(OH) were 
mapped at 110~GHz. We made maps of the integrated intensity over the velocity
range occupied by the blend of the $6_0-5_0$ 
and $6_1-5_1$ lines. The mapping was performed during excellent 
weather conditions with the system noise temperature about 350~K. 
The images were reconstructed using the maximum entropy (ME)
technique. The method used is similar to that described by 
Wilczek \& Drapatz (1985) except the computational algorithm to solve 
the optimization problem. The evolutionary algorithm, used here, 
is described in Promislov (1999). We estimated the effective HPBW of the 
reconstruction in a manner, similar to that used
by Moriarty-Schieven~et~al.~(1987). For each point of a map we subtracted 
the contribution from the source leaving us with the noise fluctuations only.
Then for each point we added the contribution from a ``test'' point source, 
located at the position of ME map brightness peak, convolved with the 
beam, and had flux equal to that of the real source. We applied the ME 
procedure to the maps produced in this manner and obtained the brightness 
distribution, the FWHP of which was adopted as the effective HPBW of the 
reconstruction.

\subsection*{3. Results}
\begin{figure*}
\includegraphics[30,400][100,800]{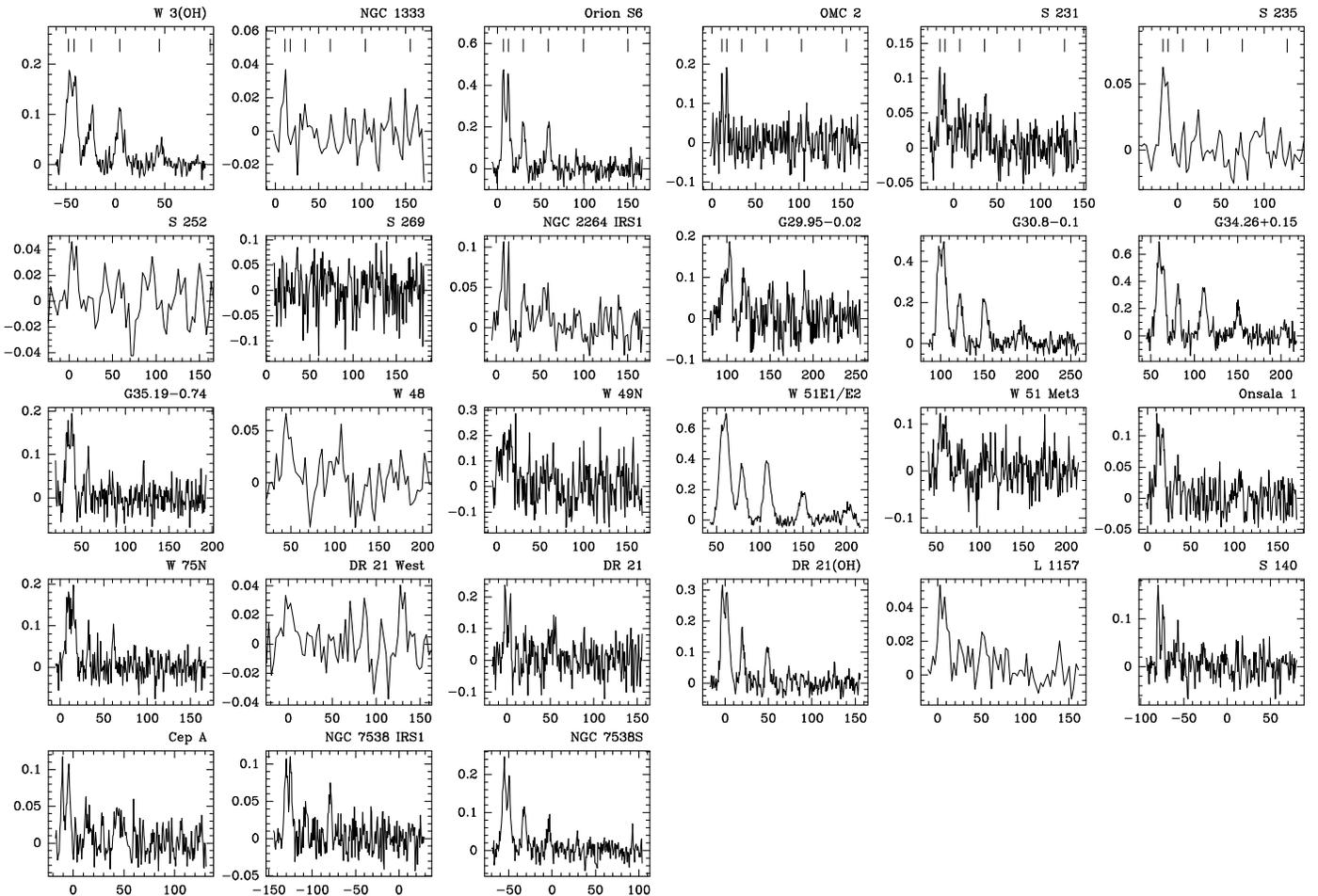}
\caption{Spectra of the sources, observed at 110 GHz. x-axis: LSR velocity of
the $6_0-5_0$ line in km~s$^{-1}$; y-axis: antenna temperature in Kelvins. 
Vertical lines in the upper row indicate the positions of 0 to 5 
$K$-components, $K$ values increase to the right.
\label{sp110}
}
\end{figure*}

\begin{figure*}
\includegraphics[30,500][100,800]{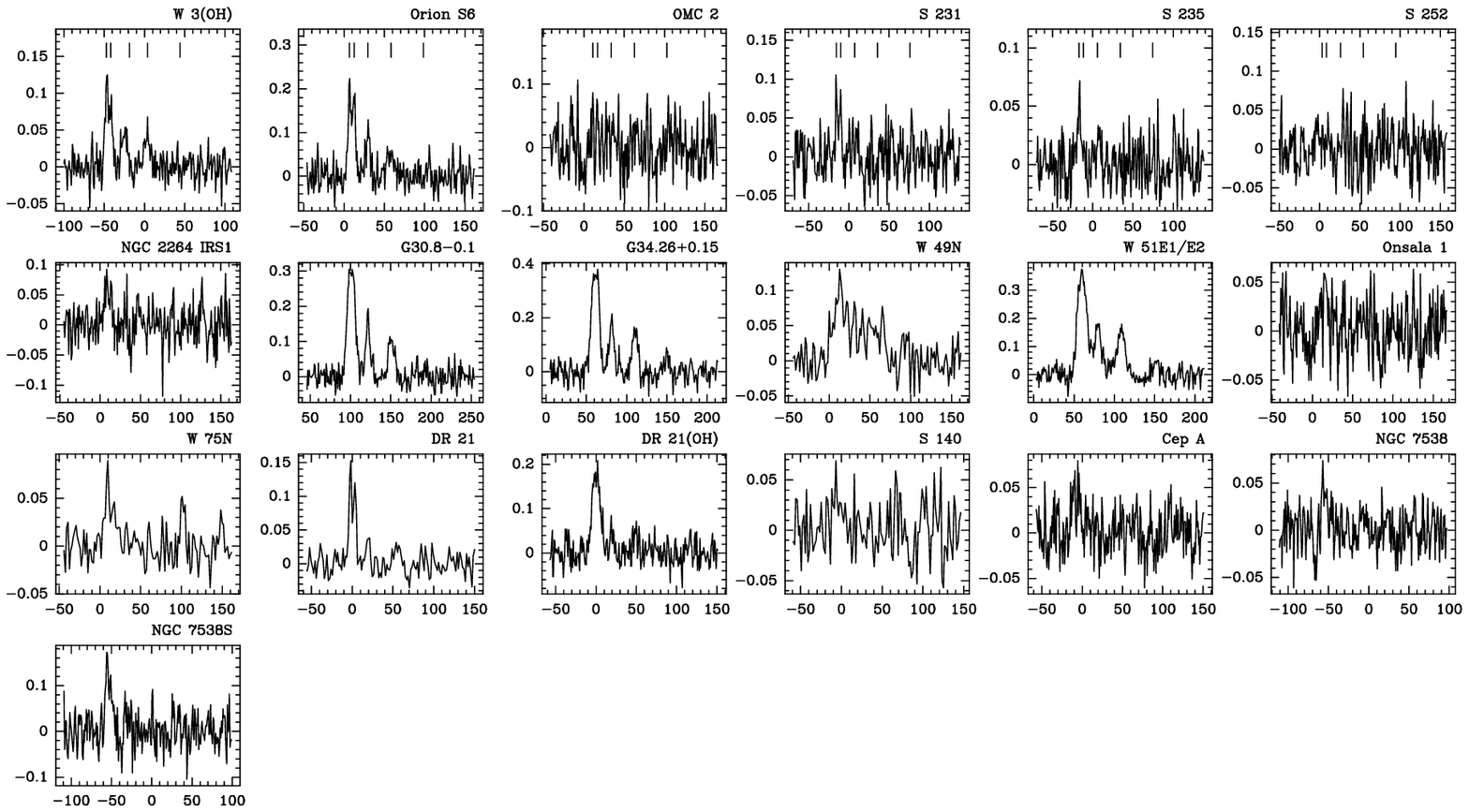}
\caption{Spectra of the sources, observed at 92 GHz. x-axis: LSR velocity of
the $5_0-4_0$ line in km~s$^{-1}$; y-axis: antenna temperature in Kelvins. 
Vertical lines in the upper row indicate the positions of 0 to 4 
$K$-components, $K$ values increase to the right.
\label{sp92}
}
\end{figure*}

\begin{figure}
\vspace{5mm}
\hskip -1cm
\resizebox{0.46\textwidth}{!}{\includegraphics{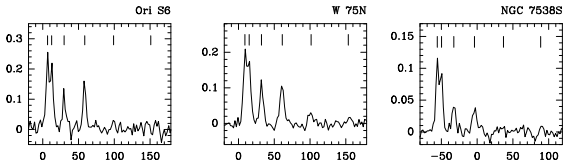}}
\vskip -12cm
\caption{Spectra of the sources, observed at 147~GHz. x-axis:  LSR velocity of
the $8_0-7_0$ line in km~s$^{-1}$; y-axis: antenna temperature in Kelvins. 
Vertical lines indicate the positions of 0 to 5
$K$-components, $K$ values increase to the right.}
\label{sp147}
\end{figure}

Emission at 110 GHz was detected towards 25 objects of the 27 observed
(S~252, W~48 and DR~21~West were only marginally detected). 
Sixteen objects were found at 92~GHz and 3 at 147~GHz. The spectra are
presented in Figs. \protect\ref{sp110}, \protect\ref{sp92}, and
\protect\ref{sp147}, respectively. 
The source coordinates and the gaussian parameters of the detected lines are 
given in Table \protect\ref{gparm}. 

\begin{table*}
\begin{center}
\scriptsize
\caption{Gaussian parameters of the detected \cyan lines with their $1\sigma$ 
errors. For each source the parameters of the 110~GHz lines are given in the 
first row and those of the 92~GHz lines in the second row. For Ori~S6,
W~75N($0'',0''$), and NGC~7538S the parameters of the 147~GHz lines are
given in the third rows.
\label{gparm}}
\tiny
\begin{tabular}{|l|rcccccccc|}
\hline\noalign{\smallskip}
Source    &R.A.$_{\rm B1950.0}$&\multicolumn{6}{c}{$\int\;T^*_{\rm A}$d$V$ (K km s$^{-1}$)}                            &$V_{\rm LSR}$&$\Delta V$\\
          &DEC.$_{\rm B1950.0}$&    $K=0$    &$K=1$     &$K=2$      &$K=3$     &$K=4$     &$K=5$   &(km s$^{-1}$)&(km s$^{-1}$)\\
\noalign{\smallskip}
\hline\noalign{\smallskip}
W 3(OH)   &02 23 17.3&0.97(0.06)&1.06(0.06)&0.80(0.01)&0.79(0.04)&0.24(0.04)& $<$0.12&$-$47.5(0.2)&8.4(0.2)\\
          &61 38 58  &0.61(0.05)&0.47(0.05)&0.29(0.04)&0.27(0.04)&$<0.09$   &          &$-$47.5(0.2)&6.9(0.3)\\
NGC 1333  &03 25 55.0&$<$0.18    &         &          &          &          &          &          &        \\
          &31 04 00  &not observed&         &          &          &          &          &          &        \\
Ori S6    &05 32 44.8&2.12(0.08)&2.00(0.08)&1.03(0.07)&0.83(0.07)&$<$0.21   &          &6.9(0.1)  &4.2(0.1)\\
          &$-$05 26 00&0.97(0.07)&0.86(0.07)&0.53(0.07)&0.31(0.07)&$<$0.3   &          &6.8(0.2)  &4.7(0.2)\\
	  &           &0.77(0.07)&0.68(0.07)&0.36(0.07)&0.48(0.07)&$<$0.21  &          &6.7(0.2)  &4.4(0.2)\\
OMC 2     &05 32 59.9&0.35(0.06)&0.44(0.07)& $<$0.18    &          &          &          &11.2(0.1) &2.1(0.2)\\
          &$-$05 11 29 & $<$ 0.11 &          &            &          &          &          &          &        \\
S 231     &05 35 51.3&0.33(0.04)&0.31(0.04)&0.18(0.04)  &0.23(0.04)& $<$ 0.12   &          &-15.9(0.2)&3.0(0.0)\\
          &35 44 16  &0.27(0.06)&0.28(0.07)&0.14(0.07)  & $<$ 0.09 &          &          &$-$15.6(0.4)&3.8(0.5)\\
S 235     &05 37 31.8&0.31(0.06)&0.26(0.05)&0.22(0.05)  & $<$0.15  &          &          &$-$17.1(0.6)&4.7(0.4)\\  
          &35 40 18  &0.22(0.04)& $<$ 0.1  &            &          &          &          &$-$17.1(0.0)&3.7(0.0)\\
S 252     &06 05 53.7&0.24(0.08)$^1$&0.18(0.08)$^1$& $<$ 0.24&     &          &          &2.9(0.9)  &5.0(0.0)\\ 
          &21 39 09  & $<$0.09  &          &            &          &          &          &          &        \\ 
S 269     &06 11 46.5&$<$0.06   &          &            &          &          &          &          &        \\ 
          &13 50 39.0&not observed&        &            &          &          &          &          &        \\ 
NGC 2264  &06 38 24.9&0.49(0.04)&0.41(0.04)&0.22(0.04)  &0.17(0.04)& $<$0.12 &          &8.5(0.2)  &4.0(0.2)\\
IRS1      &09 32 28  &0.34(0.07)&0.26(0.07)& $<$ 0.2    &          &          &          &7.8(0.5)  &4.0(0.0)\\  
G29.95-0.02&18 43 27.1&0.53(0.09)&0.73(0.09)&0.51(0.10)&0.39(0.09)&0.39(0.09)&$<$0.27&97.6(0.2) &5.5(0.4)\\
          &$-$02 42 36 &not observed&          &          &          &          &          &          &        \\ 
G30.8-0.1 &18 45 11.0&3.09(0.11)&2.44(0.10)&1.74(0.08)&1.63(0.08)&0.58(0.07)&0.21(0.07)&98.9(0.1) &7.0(0.0)\\
          &$-$01 57 57 &2.18(0.16)&1.36(0.15)&1.18(0.08)&1.10(0.09)&0.32(0.08)&          &99.0(0.2) &7.9(0.3)\\  
G34.26+0.15&18 50 46.1&3.72(0.15)&2.97(0.15)&2.20(0.14)&2.41(0.14)&1.31(0.13)&0.65(0.1)&58.7(0.1) &6.0(0.1)\\
          &01 11 12  &2.22(0.13)&2.06(0.12)&1.29(0.09)&1.28(0.10)&0.61(0.09)&          &58.5(0.2) &6.9(0.2)\\
G35.19-0.74&18 55 40.8&0.75(0.06)&0.88(0.07)&0.36(0.06)&$<$0.18     &          &          &33.7(0.2) &4.6(0.2)\\
          &01 36 30  &not observed&          &          &          &          &          &          &        \\ 
W 48      &18 59 13.8&0.32(0.08)$^1$&0.15(0.07)$^1$& $<$0.2   &          &          &          &43.9(0.6) &4.2(0.6)\\
          &01 09 20.0&not observed&          &          &          &          &          &          &        \\ 
W 49N     &19 07 49.9&1.71(0.89)&1.69(0.98)&0.81(0.32)&1.59(0.33)& $<$ 0.93   &          &9.4(1.6)  &16.0(1.3)\\ 
          &09 01 14  &1.05(0.19)&0.90(0.15)&0.99(0.15)&1.16(0.15)& $<$ 0.45   &          &9.2(1.0)  &19.8(1.1)\\
W 51E1/E2 &19 21 26.2&4.89(0.16)&4.21(0.15)&3.34(0.10)&3.76(0.10)&1.87(0.10)&0.67(0.07)&56.8(0.1) &9.5(0.1)\\
          &14 24 43  &2.86(0.11)&2.13(0.11)&1.63(0.11)&1.87(0.08)&0.72(0.08)&          &57.3(0.1) &10.8(0.2)\\
W 51 MET3 &19 21 27.5&0.65(0.14)&0.59(0.13)&0.50(0.10)&0.42(0.11)&$<$0.27   &          &55.5(0.6) &8.0(0.7)\\
          &14 23 52  &not observed&          &          &          &          &          &          &        \\ 
Onsala 1  &20 08 09.9&0.56(0.05)&0.55(0.05)&0.24(0.05)&$<$0.12    &          &          &11.9(0.2) &4.7(0.2)\\
          &31 22 42  & $<$0.18    &          &          &          &          &          &          &        \\ 
W 75N     &20 36 50.0&not observed&          &          &          &          &          &          &        \\ 
($0'',0''$)&42 26 58 &not observed&          &          &          &          &          &          &        \\ 
          &          &0.75(0.03)&0.67(0.03)&0.46(0.03)  &0.47(0.03)&0.34(0.03)&0.09(0.03)&9.1(0.1)  &5.6(0.1)\\
W 75N     &20 36 50.4&0.86(0.07)&0.84(0.07)&0.35(0.06)&0.36(0.06)&$<$0.18    &          &9.4(0.2)  &4.8(0.2)\\
($6'',25''$)&42 27 23&0.39(0.06)&0.30(0.06)& $<$ 0.18   &          &          &          &9.6(0.3)  &4.8(0.5)\\
DR 21 West&20 37 07.8&0.19(0.06)$^1$&0.16(0.08)$^1$& $<$ 0.21&     &          &          &$-$2.1(1.2) &5.0(0.0) \\ 
          &42 08 44.0&not observed&          &          &          &          &          &          &        \\ 
DR 21     &20 37 13.0&0.61(0.09)&0.47(0.08)&0.16(0.08)&0.23(0.09)  & $<$ 0.21 &          &$-$2.1(0.2) &2.7(0.0)\\
          &42 08 50.0&0.63(0.05)&0.49(0.05)&0.20(0.05)&0.11(0.05)  & $<$ 0.15 &          &$-$2.2(0.1) &3.7(0.2)\\
DR 21(OH) &20 37 13.8&1.53(0.06)&1.43(0.06)&0.74(0.05)&0.64(0.05)  & $<$0.15  &          &$-$3.1(0.1) &4.7(0.1)\\ 
          &42 12 13  &0.97(0.07)&1.03(0.07)&0.30(0.06)&0.25(0.06)  & $<$ 0.21 &          &$-$3.0(0.2) &4.7(0.0)\\
L 1157    &20 38 39.6&$<$0.09   &          &          &            &          &          &          &        \\
($0'',0''$)&67 51 33&not observed&         &          &            &          &          &          &        \\
L 1157    &20 38 41.0&0.23(0.05)&0.21(0.05)&$<$0.12   &            &          &          &2.1(0.5)  &4.9(0.6)\\
($20'',-60''$)&67 50 33  &not observed&         &         &            &          &          &          &        \\ 
S 140     &22 17 41.2&0.39(0.04)&0.30(0.04)&0.26(0.05)&$<$0.12     &          &          &$-$6.8(0.1) &3.1(0.2)\\
          &63 03 43  &0.34(0.12)&0.14(0.09)&0.15(0.09)& $<$ 0.36   &          &          &$-$7.0(0.6) &3.8(1.1)\\
Cep A     &22 54 19.2&0.30(0.04)&0.31(0.04)&0.08(0.03)&0.17(0.04)&$<$0.12    &          &$-$9.9(0.2)&3.6(0.3)\\
          &61 45 47  &0.24(0.05)&0.22(0.05)& $<$ 0.15   &          &          &          &$-$10.3(0.4)&3.9(0.5)\\ 
NGC 7538  &23 11 36.6&0.45(0.04)&0.42(0.04)&0.22(0.04)&0.25(0.04)&$<$0.12   &          &$-$57.7(0.2)&4.0(0.2)\\
IRS1      &61 11 50  &0.25(0.05)&0.15(0.05)& $<$ 0.12   &          &          &          &$-$57.3(0.5)&4.0(0.0)\\ 
NGC 7538S &23 11 36.1&1.06(0.05)&0.84(0.05)&0.57(0.05)&0.34(0.04)&$<$0.12    &          &$-$55.4(0.1)&4.5(0.1)\\
          &61 10 30  &0.74(0.09)&0.41(0.09)&0.22(0.08)& $<$ 0.24   &          &          &$-$55.6(0.4)&4.5(0.4)\\
	  &          &0.40(0.02)&0.33(0.02)&0.18(0.02)&0.16(0.02)  &$<$0.05   &          &$-$55.7(0.1)&5.1(0.2)\\
\noalign{\smallskip}
\hline
\end{tabular}\end{center}
$^1$--marginal detection
\end{table*}

\begin{figure*}
\vspace{9.2cm}
\rotatebox{-90}{\resizebox{!}{\textwidth}{\includegraphics{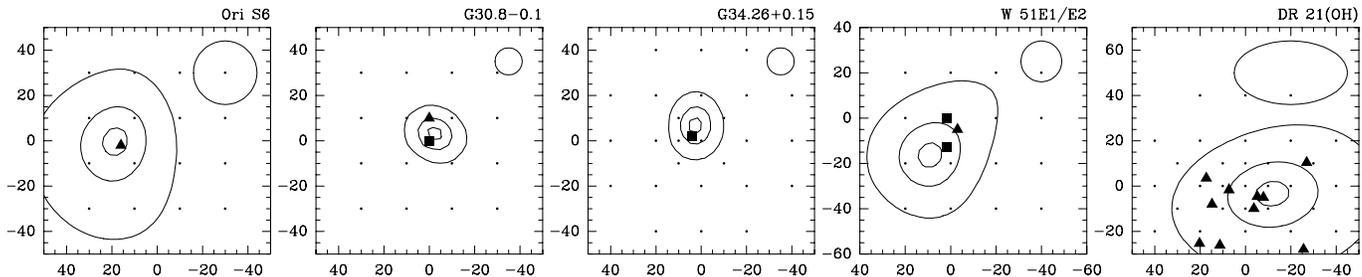}}}
\vskip -8cm
\caption{Maximum entropy maps of five sources. x and y axes are R.A. and DEC.
offsets (arcsec), respectively, from the (0,0) positions, given in Table
\protect\ref{gparm}. The observed positions are marked by dots. The contours 
correspond to 10, 50 and 90\% of the deconvolved integrated intensity peak
values, which are 21.9~K~km s$^{-1}$ for Ori~S6, 57.7~K~km s$^{-1}$ for
G30.8-0.1, 77.6~K~km s$^{-1}$ for G34.26+0.15, 25.6~K~km s$^{-1}$ for
W~51E1/E2, and 7.3~K~km s$^{-1}$ for DR~21(OH). The effective HPBW of the
reconstruction are represented by ellipses in the upper right corners. For
DR~21(OH) the resolutions in the two coordinates are different due to the
cosine effect. Filled squares show H$_2$O maser positions, filled triangles
show the positions of thermal methanol peaks, taken from Menten~et~al.~1988
(Ori~S6), Liechti \& Wilson~1996 (G30.8-0.1, G34.26+0.15, W~51E1/E2);
Liechti \& Walmsley~1997 (DR~21(OH)).
\label{maps}}
\end{figure*}

The ME images are shown in Fig.~\protect\ref{maps}. Due to a number
of reasons (mainly, relatively large telescope beam and not
sufficient sampling and sensitivity) we could not resolve the sources
even with the ME technique. On our maps the FWHP of the image
of each source proved to be very close to the corresponding effective
HPBW of the reconstruction. Thus, the results of the mapping are consistent
with compact source structures, supporting the indirect size estimates,
given in Table \ref{sparm}.

Besides the main \cyan isotopomer lines, a weak blend of the $5_0-4_0$ and
$5_1-4_1$ CH$_3^{13}$CN lines at 91941.596 and 91939.834 MHz,
respectively, was detected towards G34.26+0.15. The integrated intensities
of the $K=0$ and $K=1$ lines are 0.27 and 0.31~K~km~s$^{-1}$. The ratios
of the CH$_3^{13}$CN/CH$_3^{12}$CN line intensities, about 0.15, are much
higher than the abundance of $^{13}$C relative to $^{12}$C,
strongly suggesting that the \cyan lines are optically thick. This
assumption was confirmed by rotational diagrams and statistical equilibrium 
(SE) calculations (see below). Similar ratios between the 
$6_K-5_K$ CH$_3^{13}$CN/CH$_3^{12}$CN line intensities in G34.26+0.15 were 
found by Akeson \& Carlstrom (1996).

In all the observed sources we detected the 1--0 $^{13}$CO line. In
addition, in several objects we found HNCO, HCOOCH$_3$, and the deuturated
species NH$_2$D and CH$_3$OD. These data will be discussed elsewhere.

\subsection*{4. Rotational diagrams}

\begin{figure}
\small
\resizebox{0.46\textwidth}{!}{\includegraphics{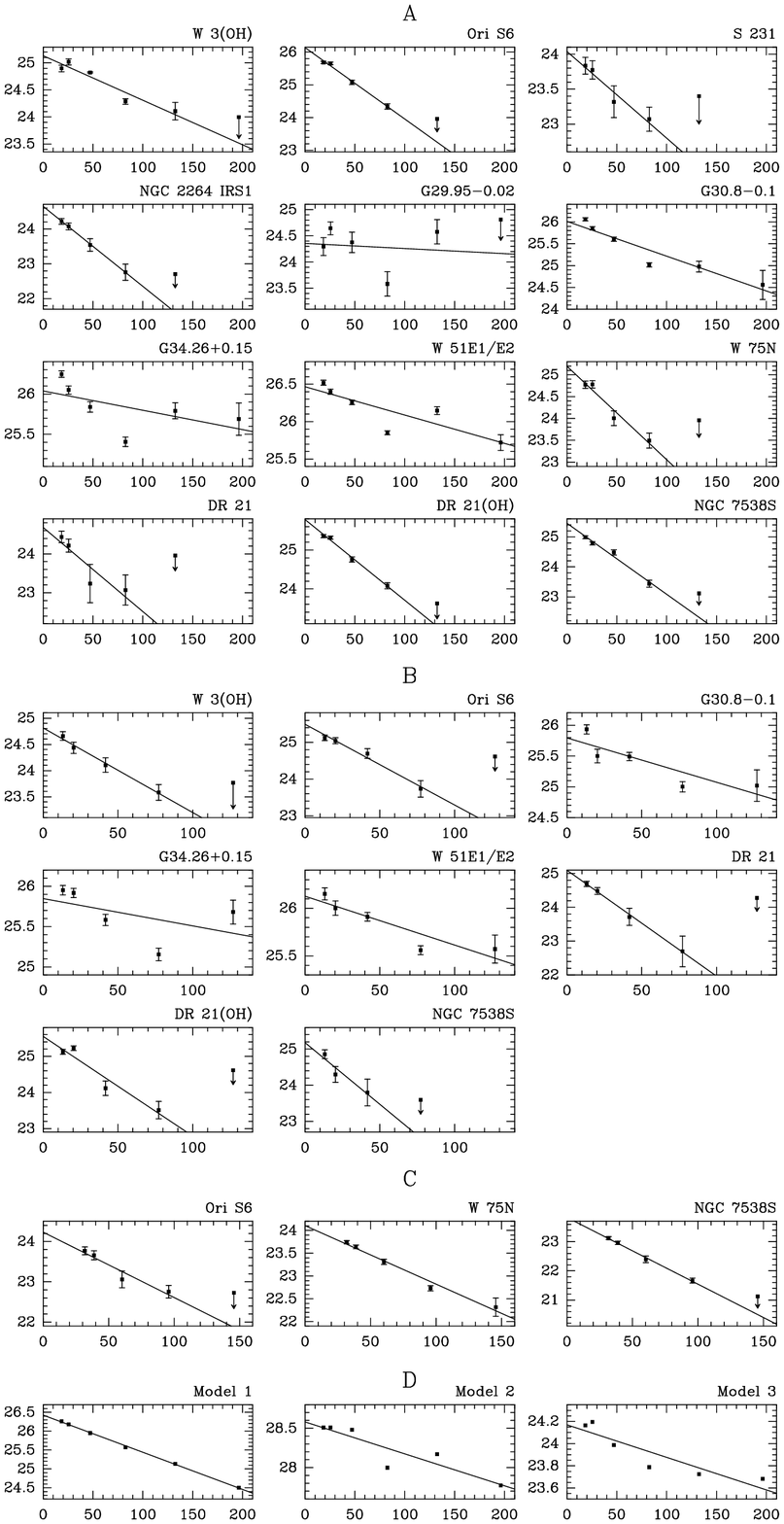}}
\caption{\small Rotational diagrams: $\ln (N_{\rm u}/g_{\rm u})$ versus
$E_u/k$~(K).
Arrows denote upper limits at 3$\sigma$ level.
{\bf A,}~110 GHz; {\bf B,}~92~GHz; {\bf C,}~147~GHz;
{\bf D,}~model rotational diagrams for the 110~GHz lines. 
The model parameters are as follows: Model~1, $n_{\rm H_2}$=10$^6$~cm$^{-3}$,
$T_{\rm kin}=100$~K, optical depth of the $6_3-5_3$ line $\tau_3=0.1$;
Model~2, $n_{\rm H_2}$=10$^6$~cm$^{-3}$,
$T_{\rm kin}=100$~K, $\tau_3=1.8$;
Model~3, $n_{\rm H_2}$=$3\times~10^3$ cm$^{-3}$,
$T_{\rm kin}=100$~K, $\tau_3=3.5$.
\label{rot}
}

\end{figure}

The molecular gas properties were derived in two ways. First, the rotational 
diagram analysis, as described by Turner (1991) was carried out. The Boltzmann
plots are shown in Fig.~\protect\ref{rot}. Since the the sources are
typically smaller than the beams and the beam filling factors at 110, 92,
and 147~GHz are different,
we analyzed different groups of lines separately. The same main features
are present both in the 110 and 92~GHz rotational diagrams.

Fig. \protect\ref{rot} shows that the majority of the plots cannot be fitted 
by straight lines. First, in some objects the points that correspond to the 
$K=3$ levels are located much lower than the fitted lines. Examination 
of previous data (e.g. line intensities, published by Churchwell et al.
1992, or Olmi et al. 1993) in a number of sources shows the same $K=3$ 
decrease, which proved to be a rather common feature and should be explained.
Symmetry requirements divide the \cyan molecule into two independent species,
$A$ and $E$, with the $K=3n$ rotational levels belonging to the $A$ species 
and the $K=3n\pm 1$ levels belonging to the $E$ species. The $K=3$ levels 
belong to the $A$ species suggesting that the $A$ species may be
underpopulated relative to $E$; but in this case the $K=0$ levels should be 
underpopulated as well, in contradiction with Fig. \protect\ref{rot}. 

Presumably the  $K=3$ decrease is caused by high optical depths. 
We constructed rotational diagrams following Turner (1991), i.e. we plotted
$N_{\rm u}/g_{\rm u}$ (see Eq. \protect\ref{eq1}) versus $E_u/k$. 
Here $g_{\rm u}$ is the upper level statistical weight, equal to the 
product of the rotational degeneracy $g_{\rm J}=(2J+1)$, the $K$-level
degeneracy $g_{\rm K}$, and the reduced nuclear spin statistical weight
$g_{\rm I}$. In symmetric tops the product $g_{\rm I} g_{\rm K}$
for the $K=3n,\;\; n\ne 0$ levels is two times larger than that for
the other levels (Turner 1991). The level populations for rotational 
diagrams are calculated assuming optically thin lines
\begin{equation}
\frac{N_{\rm u}}{g_{\rm u}}=\frac{3k\; \int T_{\rm br}\,dV}{8\pi^3 \mu^2 \nu S_{\rm ul} g_{\rm I} g_{\rm K}}
\label{eq1}
\end{equation}
Here $\mu$ is the permanent dipole moment, $\nu$ is the line frequency,
$S_{\rm ul}$ is the line strengh, $k$ is the Boltzmann constant and
$T_{\rm br}$ is the brightness temperature.
We used the main-beam brightness temperature $T_{\rm mb}$ instead of the
unknown $T_{\rm br}$. Eq. \protect\ref{eq1} implies that $\int T_{\rm br}\;dV$
is proportional to optical depth. In the case of optically thick emission 
Eq.~(\protect\ref{eq1}) underestimates $N_{\rm u}/g_{\rm u}$, the higher 
the optical depth (proportional to the statistical weight of the upper 
level), the stronger is the underestimate. 
So, the population of the more degenerate $K=3$ levels is 
underestimated to a larger degree than the population of the 
adjacent levels.

\begin{table}
\caption{Parameters derived from the rotational diagram analysis.
Columns: 1, source name; 2, group (see text for definition); 3, rotational
temperatures derived from the~110~GHz lines; 4, rotational temperatures
derived from the~92~GHz lines; 5, \cyan column densities derived from the~110~GHz lines.
\label{trot}}
\footnotesize
\tabcolsep=4pt
\medskip
\begin{tabular}{lccccc}
\hline\noalign{\smallskip}
Source     & Group&\multicolumn{3}{c}{$T_{\rm rot}$ (K)} & $N_{\rm CH_3CN}$\\
           &      & (110)            & (92)       & (147)&($10^{12}$cm$^{-2}$)\\
\noalign{\smallskip}
\hline\noalign{\smallskip}
W~3(OH)    & I$^1$&120               & 63         &      &  \\
Ori~S6     & II   &46                & 46         & 62   &33\\
S 231      & II   &80                &            &      &9 \\
NGC 2264   & II   &43                &            &      &7 \\
G29.95-0.02& I    &1013              &            &      &  \\
G30.8-0.1  & I    &125               & 142        &      &  \\
G34.26+0.15& I    &418               & 295        &      &  \\
G35.19-0.74& II   &40                &            &      &11\\
W 49N      & II   &117               &            &      &76\\
W 51E1/E2  & I    &265               & 349        &      &  \\
W 51 Met3  & II   &77                &            &      &18\\
Onsala 1   & II   &36                &            &      &7 \\
W 75N($0'',0''$)&II&                 &            & 78   &  \\
W 75N($6'',25''$)&II&47              &            &      &13\\
DR 21      & II   &23                & 32         &      &5 \\
DR 21(OH)  & II   &48                & 36         &      &24\\
S 140      & II   &42                &            &      &18\\
Cep A      & II   &57                &            &      &5 \\
NGC 7538   & II   &62                &            &      &9 \\
NGC 7538S  & II   &42                & 30         & 44   &15\\
\noalign{\smallskip}
\hline\noalign{\smallskip}
\end{tabular}
$^1$~W~3(OH) is referred to the group I since the $K=3$ decrease
is present at 110~GHz and the nature of this source (hot core) is similar
to that of other group I sources (see below).
\end{table}

The effect of optical depth is demonstrated in Fig. \protect\ref{rot}(D)
where model rotational diagrams for three parameter sets are shown. One
can see in the diagram for high density and optically thick lines the same
$K=3$ decrease. Thus, the rotational diagrams, like the marginal detection of
the $^{13}$\cyan lines, suggest that the \cyan lines may be optically thick.

Fig. \protect\ref{rot}(D) shows that for a density of $3\times 10^3$
cm$^{-3}$ there is no $K=3$ decrease even in the case of optically thick
lines. For low density and optically thick emission the excitation of the
$5_K-4_K$ and $6_K-5_K$ lines is determined by internal radiation rather
than collisions. Larger optical depth of the $K=3$ lines leads to higher
excitation and brightness temperature of these transitions and to the
absence of
the $K=3$ decrease. Examinaton of a large number of models showed
that a significant $K=3$ decrease appears if the density is of the order
of $10^5$ cm$^{-3}$ or larger.
Hence, the absence of the $K=3$ decrease shows that either
the lines are optically thin or the gas density is low.

Rotational temperature, derived from optically thick lines, is larger
than gas kinetic temperature. For example, the rotational temperature for
model 2 is 246~K and that for model 3 is 341~K versus the gas kinetic 
temperature 100~K.

The parameters, derived from the rotational diagrams are presented in
Table~\protect\ref{trot}.
On the basis of the discussion above the sources were divided into
two groups. The sources in group I are those that show significant
$K=3$ decrease. We can conclude that the 110 and 92 GHz lines are optically
thick and that the gas density is about $10^5$~cm$^{-3}$ or larger.
Since for optically thick $J_K-(J-1)_K$ \cyan lines rotational diagrams
give exaggerated temperatures, the kinetic temperatures of
group~I sources in Table~\ref{trot} are upper limits.
On the other hand, high optical depth, as well as beam dilution (note,
that both ME images and indirect size estimates in Table~\ref{sparm}
demonstrate that the group~I sources are very compact), leads to an
underestimation of column density. Hence, the column
densities of group~I sources in Table~\ref{trot} are lower limits.
The sources in group II are those where the $K=3$ decrease is either
absent or cannot be found since the $K = 4$ lines, adjacent to the $K = 3$ 
lines, are not detected.
The reliability of the
rotational temperatures and column densities of the group~II sources is
discussed in section~7. 

In G34.26+0.15, and, probably, G30.8-0.1 and W~51E1/E2 the low
excitation $K=0$ and $K=1$ levels have excess population. The same effect can
be noticed in the model rotational diagram for low density 
[Fig.~\protect\ref{rot}(C)]. However, the excess population is present in 
the rotational diagrams of group~I sources, which we believe to
be dense. Therefore this excess indicates their complex structure. 
For simplicity we call them core--halo objects. Here "halo" means any colder 
or less dense source(s) than "core".
Core-halo structure of \cyan emission has been already detected in
several objects (Olmi et al.~1996).

\subsection*{5. Statistical equilibrium calculations}

To determine source properties by means of statistical equilibrium (SE)
calculations, we used the approach, similar to that applied to \cyan 
analysis, e.g. by Bergman \& Hjalmarson (1989) and Olmi et al. (1993). The 
LVG code was kindly made available by C.M.~Walmsley and R.~Cesaroni. The free
parameters of the model are the gas kinetic temperature $T_{\rm kin}$, the
molecular hydrogen number density $n_{\rm H_2}$, and the methyl cyanide
density, divided by velocity gradient $n_{\rm CH_3CN}/(dV/dR)$.

To obtain SE parameters, one should perform SE calculations for a number 
of parameter sets and choose the sets which are in  agreement with 
observational data. The agreement between our observations and SE models 
can be evaluated by comparing model line intensity ratios within the 110,
92, and 147~GHz series with the corresponding observed ratios. For each
line series we used the ratios
$R_{\rm i}=T_{\rm mb}(J_K-(J-1)_K)/T_{\rm mb}(J_2-(J-1)_2)$.
 
We modelled the \cyan brightness temperatures for the kinetic temperatures 
10--500~K, the molecular hydrogen number densities 
$3\times 10^3$--$10^8$ cm$^{-3}$, and the \cyan densities, divided by a
velocity gradient in the range $10^{-7}$--$10$~cm$^{-3}$/(km s$^{-1}$~pc$^{-1}$).
External radiation, except the microwave background, was neglected. The 
best-fit models are those that minimize the $\chi^2$ value
\begin{equation}
\chi^2=\sum_i \left(\frac{R_{\rm i}^{\rm obs}-R_{\rm i}^{\rm mod}}
{\sigma_{\rm i}^{\rm obs}}\right)^2
\label{eq3}
\end{equation}
where $\sigma_{\rm i}^{\rm obs}$ are the rms errors of the observed ratios. 
We calculated $\chi^2$ for all models and then found the minimum values.
The 1$\sigma$ confidence levels were calculated as described in
Lampton~et~al.~(1976).

For G30.8-0.1, G34.26+0.15 and W~51E1/E2 the minimum $\chi^2$ values are
much larger than expected from the number of degrees of freedom. This fact 
confirms our suggestion that these sources cannot be described within the
frames of a homogeneous source model.
Bearing in mind our assumption about their core--halo structure
and arbitrarily suggesting that the core contribution dominates
in the $K\ge 2$ emission, we 
determined the core parameters using only the $K\ge 2$ lines (core parameters
below). This procedure virtually did not change temperatures
derived from all the observed lines. For 34.26+0.15 an interferometric 
hot core spectrum at 110 GHz (Akeson \& Carlstrom, 1996) is well approximated 
by our core models, showing that this approach is reasonable.

Finally, for each model, appropriate for a given source, we estimated the 
source size $\theta_{\rm s}$, assuming a Gaussian brightness distribution
\begin{equation}
\frac{\theta^2_{\rm s}}{\theta^2_{\rm s}+\theta^2_{\rm mb}}=\frac{T_{\rm mb}}{T_{\rm mod}}
\end{equation}
where $T_{\rm mod}$ is the model brightness temperature and 
$\theta_{\rm mb}$ is the FWHP.  
Using published distances, we estimated source linear sizes $D$, averaged 
velocity gradients $\Delta V/D$ and \cyan abundance 
\begin{equation}
X_{\rm CH_3CN}=n_{\rm m}\;\Delta V/(D\;n_{H_2})
\end{equation}
where $n_{\rm m}$ is the model $n_{\rm CH_3CN}/(dV/dR)$ value.

Like previous workers in the field, we found that only the kinetic temperature
can be determined accurately enough by this method. For most sources,
especially those of the group II, we could not obtain any density and hence,
\cyan abundance
estimate. To constrain these parameters, we left for further analysis only
those models that agree with the observed ratios between the 110, 92 and
147~GHz line intensities, assuming that the observed intensities are
accurate within 20\%.

Our SE calculations confirm that the $K=3$ decrease in the rotational
diagrams appears as a result of line opacities. For the sources, which were
considered optically thick due to the $K=3$ decrease we found that only
optically thick models correspond to the observational data.
Model gas temperatures appeared
lower than rotational temperatures for the same sources, as it is expected
for optically thick lines. For other objects, the kinetic temperatures proved
to be close to the rotational temperatures, and either only optically thin or
both moderately optically thick and optically thin models satisfy the
observational data.

The results of the SE analysis are presented in Table~\protect\ref{sparm}.
For group I objects the SE modelling shows that only very compact sources
satisfy the observed line intensity ratios. This conclusion is supported
by the maximum entropy maps.

\begin{table*}
\caption{\small Properties of the sources obtained from SE calculations. Columns:
1, source name; 2, kinetic temperature; 3, \cyan abundance; 4, adopted 
distance; 5, angular diameter; 6, density.
\label{sparm}}
\smallskip
\small
\tabcolsep=20pt
\begin{tabular}{lrrrrr}
\hline\noalign{\medskip}
Source  & $T_{\rm kin}\quad $ &$X_{\rm CH_3CN}$&$d\;\;$ &$\theta\quad$&$n_{\rm H_2}\quad$\\
        &${\rm (K)}\quad $  &$(\times 10^{-10})$&${\rm (kpc)}$&${\rm ('')}\quad$&$(10^4 {\rm cm}^{-3})$\\
\noalign{\smallskip}
\hline\noalign{\smallskip}
W 3(OH)     & 70(50--95)   &10--1000    & 2.2       & 2-4$^b$& $\ge$10 \\
Ori S6      & 45(40--50)   &0.3--30     & 0.5       &$\ge 10$& 10--100\\
S 231       & 55(35--160)  &            & 2         &$\ge 3$ &        \\
NGC 2264    & 45(25--65)   &            & 0.8       &$\ge 3$ &        \\
G30.8-0.1$^a$&60(45--75)   &$\ge 100$   & 8         & 2-7    &        \\
G34.26+0.15$^a$&140(65--220)&$\ge 300$  & 3.8       & 1-6$^c$&        \\
W 51E1/E2$^a$&160(120--180)&100--1000   & 8         & 1-3    &$\ge 30$\\
W 75N$(6'',25'')$&40(35--60)&           & 3         &$\ge 5$ &$\ge 1$ \\
DR 21       & 30(20--55)   &            & 3         &$\ge 5$ &        \\
DR 21(OH)   & 45(40--55)   &1--100      & 3         &$\ge 8$ &$\ge 3$ \\
NGC 7538    & 50(35--100)  &            & 3.5       &$\ge 3$ &        \\
NGC 7538S   & 40(30--45)   &10--100     & 3.5       & 5-9    & 10--100\\
\noalign{\smallskip}
\hline\noalign{\smallskip}
\end{tabular}

$^a$--core parameters\\
$^b$--less than $1''$ according to Wink et al. (1994).\\
$^c$--$3.9'' \times 3.3''$ (Akeson \& Carlstrom, 1996).\\
\end{table*}

\subsection*{6. Comments on selected sources}
\noindent
{\it W 3(OH).}  Wink et al. (1994) published a map of this source in the 
$5_3-4_3$ \cyan line, obtained with the Plateau de Bure
interferometer and found two compact objects, a strong clump located
toward the water maser cluster and HCN peak, known as TW object (Turner
\& Welch, 1984), and a much weaker one toward the UC HII region. Our 
temperature estimate is in a good agreement with that of Wink et al.;
however, we found that source size is larger than $2''$, in conflict
with their upper limit of $1''$ for the stronger clump. For
the core models the lower limit appeared to be less than $1''$. We suggest
that our single-dish CH$_3$CN line ratios are affected by the contribution
from the weaker clump and by a possible halo emission,
which is too weak or too extended to be found with the
Plateau de Bure interferometer.

\noindent
{\it G34.26+0.15.} An interferometer map of this well-known hot core
(Ohishi 1996) in the $6_K-5_K$ \cyan lines have been already obtained by
Akeson \& Carlstrom (1996). The ratios of all the observed by us line
intensities toward this source could not be fitted by homogeneous source
models, suggesting a core--halo structure. The core model, which is in the
best agreement with our data, fairly well represents the interferometric
spectrum by Akeson \& Carlstrom. The core temperature appeared to be about
140~K, although with large uncertainity. This value is much lower than the
250~K obtained by Akeson \& Carlstrom; the discrepancy is probably appeared
due to different accounts of optical depth effects. We could not obtain
any density estimate from our data only;
however, the brightness temperature, which can be obtained using the
source size by Akeson \& Carlstrom (1996) requires a density larger than 
$10^5$ cm$^{-3}$.

\noindent
{\it W 51E1/E2.} The \cyan source was found toward the radio continuum
objects E1 and E2, which are associated with the H$_2$O and OH masers W~51
Main/South, ammonia peaks NH$_3$-1 and NH$_3$-2, detected at VLA
(Ho~et~al.~1983), and emission of other molecules, typical for hot cores
(Ohishi~1996). The peak of the \cyan emission is coincident within the pointing
error of $5''$ with the ammonia clump NH$_3$-1; however, the peak position
accuracy is not good enough since the peak is close to the edge of the
mapped region. Without further observations we cannot state that the \cyan
peak is coincident with any of the mentioned objects. There is a weak tongue
towards the peak of thermal methanol emission (Liechti \& Wilson 1996),
suggesting that the structure of the \cyan source is complex. The observed 
lines cannot be fitted within the frames of the model of a homogeneous 
source. The core models give a kinetic temperature of about 160 K, density 
larger than $3\times 10^5$ cm$^{-3}$, and core diameter between $1''$ and 
$3''$.

\noindent
{\it DR 21(OH).} The peak of the \cyan brightness coincides with the methanol
maser and peak of thermal emission DR 21(OH)-2 (Liechti \& Walmsley 1997).
We could not find a homogeneous source model that fits the observed data
well enough ($\chi^2_{\rm min}=8.4$ instead of expected value of 3),
suggesting that the structure is complex, as it was already found for
methanol and other molecules (Liechti \& Walmsley 1997 and references
therein).  However, the non-detection of the lines with $K > 3$ precludes
core--halo analysis.

\subsection*{7. Discussion}

SE analysis shows that the group I sources are very different from the others.
These are the strongest sources in our sample and have kinetic temperatures
higher than the group II sources. All of them, except
\mbox{G30.8-0.1}, are well-known hot cores (Wink et al. 1984; Ohishi 1996). The
\cyan abundance in these objects is about or larger than
10$^{-9}$ and larger than in the group~II objects. The \cyan
abundance of order of 10$^{-8}$ have been reported toward hot cores, 
e.g. in Orion (Blake et al. 1987) and in G10.47+0.03 (Olmi et al. 1993).
Presently, the fact that the \cyan abundance in hot cores is increased due to
evaporation of molecular material from icy mantles of interstellar grains is
well established. \cyan may be either a parent molecule or may be a product
of a chain of gas-phase reactions, starting from HCN or some other
nitrogen-bearing molecule, which appears in gas phase due to mantle
evaporation (see e.g., Millar 1996 and references therein). This scenario
is in agreement with the current observational data on hot cores and has
further support in the results of the observations of comets. Both HCN and
\cyan, as well as some other nitrogen-bearing species, were identified as 
parent molecules in cometary comae (Bockel\'ee-Morvan 1996; Biver et al.
1998). Since cometary ices may consist of interstellar material, this is an
indication that grain mantle evaporation can be a source of \cyan.

Most hot cores, observed in \cyan lines, are sites of massive star formation.
Other regions, where gas-phase chemistry is strongly affected by grain
mantle evaporation, are bow shocks at the edges of y oung bipolar outflows,
driven by low-to-intermediate mass protostars (Bachiller 1996). The temperature
in these objects is high enough to evaporate grain mantles, and ammonia and
HCN abundance is strongly enhanced here, as well as the contents of some
other molecules, abundant in hot cores. Thus, one can expect high \cyan
abundance in these regions. We observed two such objects, NGC~1333~IRAS2 and
L~1157. In L~1157, no lines were found toward the central source
($0'',0''$ position), but a weak emission was detected toward the blue wing
of the outflow ($20'',-60''$ position), where the abundance of
HCN, NH$_3$, CH$_3$OH and some other molecules is enhanced up to several
orders of magnitude (Bachiller \& P\'erez Guti\'errez 1997, and references
therein). We estimated the \cyan abundance assuming that the \cyan emission
arise in the same small region as the NH$_3$ emission (B1 in the notation of
Tafalla \& Bachiller 1995). Assuming B1 size about $10''$--$15''$ and
kinetic temperature 80~K (Tafalla \& Bachiller 1995) we obtained the \cyan
column density of about $2\times 10^{13}$ cm$^{-2}$. H$_2$ column density of
$2.4\times 10^{21}$ cm$^{-2}$ follows from the $^{18}$CO data by
Bachiller \& P\'erez Guti\'errez (1997). Hence, \cyan abundance is about
$10^{-8}$, as high as in hot cores.
Thus, our data show that \cyan may be enhanced toward the blue wing
of this bipolar outflow. Sensitive high-resolution observations 
are desirable to check this conclusion.

We did not detect any emission toward the red wing of the NGC~1333~IRAS2 E--W 
outflow, where CS and \meth enhancement has been found by Sandell et al. (1994),
except a weak spike at about 2.5 $\sigma$ level, which can be attributed
to the $6_0-5_0$ line. The CS column density at this position, derived by
Langer et al. (1996), is an order of magnitude lower than the CS column
density toward L~1157~B1, presented by Bachiller \& P\'erez Guti\'errez (1997),
showing that either gas column density or molecular enhancement or both 
are smaller in NGC~1333~IRAS2 E--W, leading to the non-detection of the \cyan
emission.

The nature of the group II sources is unclear. On the one hand, their
rotational diagrams do not demonstrate significant point deviations
from the fitted lines. Therefore the conditions, necessary for the
application of rotational diagrams may be fulfilled, in particular,
the sources may be uniform and the observed lines may be optically thin.
In this case, these objects are warm clouds (50~K or lower).
On the other hand, the good appearance of rotational diagrams does not
{\it prove} that these conditions are fulfilled.
The sources may be inhomogeneous or/and the lines may be
optically thick. In particular, since the intensities of
the $K = 4$ lines are unknown, we cannot distinguish between an opaque
core--halo source or an optically thin uniform cloud, colder than the core.
This statement can be explained as follows. Let us consider
a rotational diagram of an optically thick core--halo object.
A high optical depth leads to an
apparent decrease of the $N_3/g_3$ value and to the $K=3$ decrease,
while the halo contribution leads to an increase of the $N_0/g_0$ and
$N_1/g_1$ values. Therefore the slope of the fitted line, drawn over the
$K=0$--3 points, is larger than the slope that corresponds to the core,
and provides a lower temperature. Since signal-to-noise
ratios are typically rather poor in our spectra of group~II sources, neither
rotational diagrams nor $\chi^2$ analysis make it possible to distinguish
between the two source models. Therefore it is possible that at least some of
the group~II objects are the same
as those of group~I, i.e. consist of compact opaque hot cores and
colder halos.  Note, that the presence of hot gas in most of the observed
group~II sources is indicated by e.g. water vapor maser emission.
Therefore more sensitive observations are required to study the structure
of these objects. Below we
assume that the group~II sources are homogeneous; all the results, obtained
from the rotational diagrams and SE analysis are valid only if this assumption
is correct.
 
The temperatures of the group II sources proved to be about 50~K or lower
and the \cyan abundance proved to be either below $\approx 10^{-9}$ or
cannot be estimated from our data. We have no size estimates for these
sources, apart from very uncertain 
lower limits. Density estimates, where available, show that it is at least 
larger than $10^4$~cm$^{-3}$. One can suggest that the \cyan emission arise in
warm clouds, well-known from the observations of CS, NH$_3$ and other 
molecules. One of the best studied warm clouds is NGC~2264~IRS1.
Our temperature estimate from Table~\ref{sparm} agrees with the value
of~55~K, obtained by Schreyer~et~al.~1997 on the basis of
a careful study of this source in a large number of lines of different
molecules. Thus, the assumption that the~\cyan emission in
the group~II sources arise in warm (30--50~K), dense ($10^4$~cm$^{-3}$
or larger) gas may be correct, but to date we cannot exclude the
possibility that at least in some of these sources
the emission from hot and compact objects with enhanced 
\cyan abundance may be present (see the previous paragraph).

\subsection*{8. Summary and conclusions}

A survey of 27 galactic star-forming regions in the $6_K-5_K$, 
$5_K-4_K$, and $8_K-7_K$ \cyan lines at 110, 92, and 147~GHz, respectively,
was carried out.
Twenty-five sources were detected at 110~GHz, nineteen at 92~GHz and three
at 147~GHz.

A number of source parameters were derived using rotational diagram
analysis and statistical equilibrium calculations. As expected,
the strongest \cyan emission was found towards hot cores. The kinetic
temperatures of these sources are
of the order of or larger than 100~K, the \cyan abundances are larger than
$10^{-9}$, and the lines are optically thick.
Hot cores are usually associated with colder/less dense
gas. Probably, the~\cyan abundance
in the hot gas, associated with the young bipolar outflow L~1157, is 
about $10^{-8}$. 

The nature of weaker objects is unclear. \cyan emission may arise in warm
(30--50~K) dense ($>10^4$ cm$^{-3}$) clouds. However, at least in some of
these sources hot compact objects with enhanced \cyan abundance may be
present.\\[10pt]
{\footnotesize\it Acknowledgements}
\footnotesize The authors are grateful to the staff of the Onsala Space 
Observatory for providing help during the observations. We are grateful
to Dr. S. Radford (NRAO) for scheduling the 147~GHz observations and the staff
of 12-m NRAO telescope for the help during the observations. We would like to
thank Dr. V.I.~Slysh for helpful discussions and Drs. C.M.~Walmsley and
R.~Cesaroni for making available the LVG code. The work was done under partial
financial support of the Russian Foundation for Basic Research (grant No
95-02-05826) and the project $N$315 ``Radio Astronomy Educational and 
Scientific Center'' within the frames of the program ``State Support for the
Integration of High School and Basic Research''. Onsala Space Observatory 
is the Swedish National Facility for Radio Astronomy and is operated by 
Chalmers University of Technology,
G\"oteborg, Sweden, with financial support from the Swedish Natural Science
Research Council and the Swedish Board for Technical Development.

\end{document}